\documentclass[pra,aps,twocolumn,floatfix,showpacs]{revtex4}
\usepackage{graphics}
\begin{document}
\title{Vortex core states in superfluid Fermi-Fermi mixtures with unequal masses}
\author{M. Iskin}
\affiliation{Joint Quantum Institute, National Institute of Standards and Technology, and 
University of Maryland, Gaithersburg, Maryland 20899-8423, USA.}
\date{\today}

\begin{abstract}

We analyze the vortex core states of two-species (mass imbalanced) superfluid 
fermion mixtures as a function of two-body binding energy in two dimensions. 
In particular we solve the Bogoliubov-de Gennes equations for a 
population balanced mixture of $^{6}$Li and $^{40}$K atoms at 
zero temperature. We find that the vortex core is mostly occupied by the light-mass 
($^{6}$Li) fermions and that the core density of the heavy-mass ($^{40}$K) fermions 
is highly depleted. This is in contrast with the one-species (mass balanced) 
mixtures with balanced populations where an equal amount of density depletion 
is found at the vortex core for both pseudospin components.

\end{abstract}

\maketitle

Recent observation of quantized vortices in two-component mass and population
balanced fermion mixtures of $^6$Li atoms with short-ranged attractive 
interactions~\cite{mit-vortex} has not only complemented the previously found 
evidence~\cite{djin, jthomas, rgrimm, csalomon, rhulet}, but also indicated 
a very strong evidence for the superfluid ground state. In these experiments
it has been realized that the ground state evolves smoothly from a 
Bardeen-Cooper-Schrieffer (BCS) superfluid to a molecular Bose-Einstein 
condensate (BEC) as the attractive interaction strength varies from small to 
large values, marking the first demonstration of the theoretically predicted BCS-BEC 
crossover~\cite{leggett, nsr, carlos}. More recently quantized vortices have 
also been studied in mass balanced but population imbalanced fermion 
mixtures~\cite{mit-mixture} to investigate more exotic superfluid phases.
On the theoretical side vortex core states in superfluid fermion mixtures 
have been extensively studied for mass and population balanced 
mixtures~\cite{gygi, nygaard, bulgac, machida, sensarma}. In these 
works it was predicted that the local density of fermions is slightly depleted
at the vortex core for weak interactions, and that the density depletion
increases with increasing interaction strength. Furthermore vortex core 
states of mass balanced but population imbalanced fermion mixtures
have recently been analyzed~\cite{takahashi, hu}, showing that the unpaired 
(excess) fermions occupy the core states. However these predictions have 
not been experimentally observed since a probing technique analogous to the
scanning tunneling microscopy is still lacking.

Arguably one of the current frontiers of ultracold atom research is the 
investigation of two-species mass imbalanced fermion mixtures (e.g. $^6$Li 
and $^{40}$K, $^6$Li and $^{87}$Sr, or $^{40}$K and $^{87}$Sr) with or 
without a population imbalance, due to their greater potential for finding 
exotic phases~\cite{taglieber, ville}. Such mixtures are currently 
of interest to many communities ranging from atomic and molecular to condensed 
and nuclear matter physics, and recent analysis of the ground state phase 
diagram have shown quantum and topological phase 
transitions~\cite{iskin-mixture, pao-mixture, lin-mixture, parish-mixture, green, iskin-trap, orso-mixture}.
In this manuscript, motivated by the very recent experimens on $^6$Li 
and $^{40}$K mixtures~\cite{taglieber, ville}, we analyze the vortex core 
states of superfluid Fermi-Fermi mixtures with unequal masses as a function 
of two-body binding energy. In particular we use the Bogoliubov-de 
Gennes (BdG) formalism to study a population balanced mixture of $^{6}$Li 
and $^{40}$K atoms at zero temperature. We find that the vortex core is mostly 
filled with the light-mass ($^{6}$Li) fermions and that the core density 
of the heavy-mass ($^{40}$K) fermions is highly depleted. This is in contrast 
with the one-species (mass balanced) mixtures with balanced populations 
where an equal amount of density depletion is found at the vortex core for both 
pseudospin components~\cite{gygi, nygaard, bulgac, machida, sensarma}.

%
%
We achieve these results by using the following Hamiltonian density 
$
H(x) = \sum_\sigma \psi_\sigma^\dagger(x)K_\sigma(\mathbf{r})\psi_\sigma(x) - g \Psi^\dagger(x) \Psi(x),
$
which describes two-component fermion mixtures with attractive $(g > 0)$ and 
short-range interactions. Here $\hbar = k_B = 1$, and 
$\psi_{\sigma}^\dagger (x)$ and $\psi_{\sigma} (x)$ 
are the Grassmann field operators corresponding to creation and annihilation of 
pseudospin $\sigma$ fermions at position $\mathbf{r}$ and time $\tau$ $(x \equiv \{\mathbf{r}, \tau\})$. 
Furthermore
$
K_\sigma(\mathbf{r}) = -\nabla^2/(2M_\sigma) - \mu_\sigma
$
and
$
\Psi(x) = \psi_\downarrow(x) \psi_\uparrow(x),
$
where $M_\sigma$ is the mass and $\mu_\sigma$ is the chemical potential 
of $\sigma$ fermions.
In the mean-field approximation for the superfluid phase the resultant Hamiltonian 
can be diagonalized via the Bogoliubov-Valatin transformation
$
\psi_\sigma(\mathbf{r}) = \sum_n[u_{n,\sigma}(\mathbf{r}) \gamma_{n,\sigma} 
- s_\sigma v_{n,\sigma}^*(\mathbf{r}) \gamma_{n,-\sigma}^\dagger],
$
where $u_{n,\sigma}(\mathbf{r})$ and $v_{n,\sigma}(\mathbf{r})$ are the amplitudes
and $\gamma_{n,\sigma}^\dagger$ and $\gamma_{n,\sigma}$ are the operators corresponding 
to the creation and the annihilation of pseudospin $\sigma$ quasiparticles,
and $s_\uparrow = +1$ and $s_\downarrow = -1$. Then the single particle Green's 
function matrix (in Nambu pseudospin space) can be written as
\begin{eqnarray}
\label{eqn:green}
\mathbf{G}(\mathbf{r},\mathbf{r'};i\omega_\ell) = \sum_{n,\sigma}
\frac{\mathbf{\varphi_{n,\sigma}}(\mathbf{r}) \mathbf{\varphi_{n,\sigma}}^\dagger(\mathbf{r'})}
{i\omega_\ell - s_\sigma \epsilon_{n,\sigma}},
\end{eqnarray}
where $\omega_\ell = (2\ell + 1)\pi T$ is the fermionic Matsubara frequency, $\ell$ is
an integer number and $T$ is the temperature. 
Here $\mathbf{\varphi_{n,\sigma}}(\mathbf{r})$ and $\epsilon_{n,\sigma} > 0$ 
are the eigenfunctions and the eigenvalues of the BdG equations
\begin{equation}
\label{eqn:bdg}
\left[ \begin{array}{cc}
K_\uparrow(\mathbf{r}) & \Delta(\mathbf{r}) \\
\Delta^*(\mathbf{r}) & -K_\downarrow^*(\mathbf{r}) 
\end{array} \right]
\mathbf{\varphi_{n,\sigma}}(\mathbf{r}) 
= s_\sigma \epsilon_{n,\sigma} \mathbf{\varphi_{n,\sigma}} (\mathbf{r}),
\end{equation}
where $\mathbf{\varphi_{n,\sigma}} (\mathbf{r})$ is given by
$
\mathbf{\varphi_{n,\uparrow}}^\dagger (\mathbf{r}) 
= [u_{n,\uparrow}^*(\mathbf{r}), v_{n,\downarrow}^*(\mathbf{r})]
$
for the $\uparrow$ and
$
\mathbf{\varphi_{n,\downarrow}}^\dagger (\mathbf{r}) 
= [v_{n,\uparrow}(\mathbf{r}), -u_{n,\downarrow}(\mathbf{r})]
$
for the $\downarrow$ eigenvalues. Since the BdG equations are invariant under the 
transformation $v_{n,\uparrow}(\mathbf{r}) \to u_{n,\uparrow}^*(\mathbf{r})$,
$u_{n,\downarrow}(\mathbf{r}) \to -v_{n,\downarrow}^*(\mathbf{r})$ and 
$\epsilon_{n,\downarrow} \to -\epsilon_{n,\uparrow}$, it is sufficient to solve only 
for $u_n(\mathbf{r}) \equiv u_{n,\uparrow}(\mathbf{r})$, 
$v_n(\mathbf{r}) \equiv v_{n,\downarrow}(\mathbf{r})$ and $\epsilon_n \equiv \epsilon_{n,\uparrow}$
as long as we keep all of the solutions with positive and negative eigenvalues.

%
%
In Eq.~(\ref{eqn:bdg}) $\Delta(\mathbf{r})$ is the local superfluid order parameter 
defined by
$
\Delta(\mathbf{r}) = g \langle \psi_\uparrow(\mathbf{r}) \psi_\downarrow(\mathbf{r}) \rangle 
= - g T \sum_{\ell} G_{\uparrow,\downarrow}(\mathbf{r},\mathbf{r};i\omega_\ell),
$
which after evaluating the frequency sum leads to
$
\Delta(\mathbf{r}) = - g \sum_{n,\sigma} 
s_\sigma u_{n,\sigma}(\mathbf{r}) v_{n,-\sigma}^*(\mathbf{r}) f(s_\sigma\epsilon_{n,\sigma}).
$
Here $\langle ... \rangle$ is a thermal average and $f(x) = 1/[\exp(x/T) + 1]$ is the 
Fermi function. Using the symmetry of the BdG equations this equation 
can be written as
$
\Delta(\mathbf{r}) = - g \sum_{n} u_{n}(\mathbf{r}) v_{n}^*(\mathbf{r}) f(\epsilon_{n}).
$
We also relate the interaction strength $g$ to the two-body binding energy $\epsilon_b < 0$ 
of an $\uparrow$ and a $\downarrow$ fermion via the relation 
$
1/g = (1/A) \sum_{\mathbf{k}} 1/(\epsilon_{\mathbf{k},\uparrow} + \epsilon_{\mathbf{k},\downarrow} - \epsilon_b),
$
where $A$ is the area of the sample and 
$
\epsilon_{\mathbf{k},\sigma} = k^2/(2M_\sigma)
$
is the kinetic energy. This leads to
$
g = 4\pi/[M_r \ln(1-2\epsilon_c/\epsilon_b)],
$
where $M_r = 2M_\uparrow M_\downarrow/(M_\uparrow + M_\downarrow)$ is twice the 
reduced mass of an $\uparrow$ and a $\downarrow$ fermion and $\epsilon_c$ is the 
energy cutoff used in the $\mathbf{k}$-space integration. 
The order parameter equation has to be solved self-consistently with the number 
equations
$
N_\sigma = \int d\mathbf{r} n_\sigma(\mathbf{r}) 
$
to determine $\mu_\sigma$, where
$
n_\sigma(\mathbf{r}) = \langle \psi_\sigma^\dagger(\mathbf{r}) \psi_\sigma(\mathbf{r}) \rangle
= s_\sigma \lim_{\tau \to 0^+} 
T \sum_{\ell} e^{is_\sigma \omega_\ell \tau} G_{\sigma,\sigma}(\mathbf{r},\mathbf{r};i\omega_\ell)
$
is the local density of $\sigma$ fermions. After evaluating the frequency 
sum this relation leads to
$
n_\sigma(\mathbf{r}) = \sum_{n} [ 
|u_{n,\sigma}(\mathbf{r})|^2 f(\epsilon_{n,\sigma}) 
+ |v_{n,\sigma}(\mathbf{r})|^2 f(-\epsilon_{n,-\sigma})
],
$
which can be written as
$
n_\uparrow(\mathbf{r}) = \sum_{n} |u_n(\mathbf{r})|^2 f(\epsilon_n)
$
and
$
n_\downarrow(\mathbf{r}) = \sum_{n} |v_n(\mathbf{r})|^2 f(-\epsilon_n)
$
by using the symmetry of the BdG equations.
Having discussed the BdG formalism, next we analyze the self-consistency equations
for a single vortex.

%
%
In particular we consider a two-dimensional homogenous disk of radius $R$ such 
that the local order parameter can be written as
$
\Delta(\mathbf{r}) = \Delta(r) \exp(-i\kappa\theta),
$
where $\mathbf{r} = (r,\theta)$ are the polar coordinates and $\kappa$ is the vortex
winding number. This choice is due to numerical reasons and we do not expect 
any qualitative difference between our results and the three-dimensional ones.
Then the normalized wave functions are of the form
$
u_n(\mathbf{r}) = u_{n,m}(r) \exp(im\theta)/\sqrt{2\pi}
$
and
$
v_n(\mathbf{r}) = v_{n,m}(r) \exp[i(m+\kappa)\theta]/\sqrt{2\pi}
$
such that the BdG equations can be solved separately in each subspace of fixed angular 
momentum $m$~\cite{gygi}. We further project the radial wave functions
$
u_{n,m}(r) = \sum_j c_{n,j} \phi_{j,m}(r)
$
and
$
v_{n,m}(r) = \sum_j d_{n,j} \phi_{j,m+\kappa}(r)
$
onto a set of orthonormal Bessel functions
$
\phi_{j,m}(r) = \sqrt{2} J_m(\alpha_{j,m} r/R) / [R J_{m+1}(\alpha_{j,m})],
$
where the argument $\alpha_{j,m}$ is the $j$th zero of $J_m(x)$. 
This procedure reduces BdG equations given in Eq.~(\ref{eqn:bdg}) to a 
$2j_{max} \times 2j_{max}$ matrix eigenvalue problem~\cite{gygi}
\begin{eqnarray}
\label{eqn:bdg.matrix}
\sum_{j,j'} \left( \begin{array}{cc}
K_{\uparrow,m}^{j,j'} & \Delta_m^{j,j'} \\
\Delta_m^{j',j} & -K_{\downarrow,m+\kappa}^{j',j} 
\end{array} \right)
\left( \begin{array}{c}
c_{n,j'} \\
d_{n,j'} 
\end{array} \right)
= \epsilon_{n} \sum_j 
\left( \begin{array}{c}
c_{n,j} \\
d_{n,j} 
\end{array} \right),
\end{eqnarray}
if we allow $1 \le j \le j_{max}$ states. Here
$
K_{\sigma,m}^{j,j'} = [\alpha_{j,m}^2/(2M_\sigma R^2) - \mu_\sigma] \delta_{j,j'}
$
is the diagonal and
$
\Delta_m^{j,j'} = \int r dr \Delta(r) \phi_{j,m}(r) \phi_{j',m+\kappa}(r)
$
is the off-diagonal element where $\delta_{i,j}$ is the Kronecker delta. 
Furthermore the order parameter equation reduces to
\begin{equation}
\label{eqn:op}
\Delta(r) = -g \sum_{n,m,j,j'} \frac{c_{n,j} d_{n,j'}}{2\pi} \phi_{j,m}(r) \phi_{j',\widetilde{m}}(r) f(\epsilon_n),
\end{equation}
and the local density equations reduce to
\begin{eqnarray}
\label{eqn:ne.up}
n_\uparrow(r) &=& \sum_{n,m,j,j'} \frac{c_{n,j} c_{n,j'}}{2\pi} \phi_{j,m}(r) \phi_{j',m}(r) f(\epsilon_n), \\
\label{eqn:ne.down}
n_\downarrow(r) &=& \sum_{n,m,j,j'} \frac{d_{n,j} d_{n,j'}}{2\pi} \phi_{j,\widetilde{m}}(r) \phi_{j',\widetilde{m}}(r) f(-\epsilon_n),
\end{eqnarray}
where $\widetilde{m} = m+\kappa$. Notice that the total numbers of $\uparrow$ 
and $\downarrow$ fermions are given by
$
N_\uparrow = \sum_{n,m,j} c_{n,j}^2 f(\epsilon_n)
$
and
$
N_\downarrow = \sum_{n,m,j} d_{n,j}^2 f(-\epsilon_n),
$
respectively. We emphasize that these mean-field equations can be used for all
values of $g$ but they only provide a qualitative description of BCS-BEC crossover 
at zero temperature ($T = 0$) as discussed next.

\begin{figure} [htb]
\centerline{\scalebox{0.5}{\includegraphics{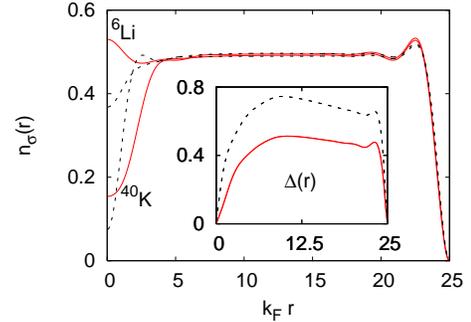} }}
\caption{\label{fig:density}
(Color online) Local density of $\sigma$ fermions $n_\sigma(r)$ [in units of $k_F^2/(2\pi)$] 
versus radius $r$ (in units of $1/k_F$) is shown for a population balanced mixture 
of $^6$Li and $^{40}$K atoms. Here $|\epsilon_b| = 0.1\epsilon_F$ 
(solid lines) and $|\epsilon_b| = 0.2\epsilon_F$ (dotted lines). The inset shows the local 
order parameter $\Delta(r)$ (in units of $\epsilon_F$) versus $r$ for the same parameters.
}
\end{figure}
%

%
%
In this manuscript we discuss a population balanced mixture of $^6$Li and $^{40}$K 
atoms, where $N_\uparrow = N_\downarrow$ and $M_\uparrow = 0.15 M_\downarrow$.
For this purpose we set a large energy cutoff $\epsilon_c = 9\epsilon_F$, and numerically 
solve the self-consistency Eqs.~(\ref{eqn:bdg.matrix}),~(\ref{eqn:op}),~(\ref{eqn:ne.up}), 
and~(\ref{eqn:ne.down}) for a singly quantized vortex with $\kappa = 1$ at $T = 0$.
Here $\epsilon_F = k_F^2/(2M_r)$ is a characteristic energy 
scale where $k_F$ is the Fermi momentum corresponding to the total density of 
fermions. We also choose $R = 25/k_F$ as the radius of the two-dimensional disc, 
and $j_{max} = 25$ and $|m|_{max} = 50$ as the maximum quantum numbers.
Notice that the bulk density $n_\sigma = N_\sigma/A$ is given by 
$n_\uparrow = n_\downarrow = k_F^2/(4\pi)$ for a population balanced mixture.
As one may expect presence of a single vortex can not significantly 
effect the bulk parameters. Therefore, to simplify the numerical calculations, 
we first solve $\mu_\sigma$ and $|\Delta_0|$ self-consistently for a
vortex-free system, and then use these solutions as an input for our 
vortex calculation. Here $|\Delta_0|$ corresponds to the bulk value 
of $\Delta(r)$. Since our vortex calculation is not fully self-consistent, 
$N_\uparrow$ and $N_\downarrow$ turn out to be very close
$|N_\uparrow - N_\downarrow| \sim 10^{-3} (N_\uparrow + N_\downarrow)$
but not exactly the same.
Furthermore, for mass balanced mixtures, we checked that this procedure 
gives results that are in qualitative agreement with the earlier works on 
population balanced~\cite{gygi, nygaard, bulgac, machida, sensarma} 
as well as population imbalanced systems~\cite{takahashi, hu}. 

In Fig.~\ref{fig:density} we show $n_\sigma(r)$ for a population balanced 
mixture of $^6$Li and $^{40}$K atoms. We also show  
$\Delta(r)$ as an inset for the same parameters.
When $g$ is small such that $|\epsilon_b| \ll \epsilon_F$, the vortex core is 
mostly filled with the light-mass ($^{6}$Li) fermions while core density 
of the heavy-mass ($^{40}$K) fermions is highly depleted. This is because the 
bound state energy spectrum is discrete with a small but nonzero separation 
$\sim |\Delta_0|^2/\epsilon_F$ (discussed below), and thus only the light-mass 
fermions can occupy these states at the core since $\mu_{Li} > \mu_{K}$ 
due to the mass difference. While this is in sharp contrast with mass and 
population balanced mixtures where the bound states
are unoccupied and equal amount of depletion occurs for both $\sigma$ 
fermions~\cite{gygi, nygaard, bulgac, machida, sensarma}, 
it is similar to mass balanced but population imbalanced mixtures where the core 
is filled with excess fermions due to their higher chemical potential~\cite{takahashi, hu}. 
However, as shown in Fig.~\ref{fig:density}, local density of the light-mass 
fermions as well as that of the heavy-mass one deplete more with increasing $|\epsilon_b|$,
which is qualitatively similar to that of mass balanced mixtures
where the local density depletion also increases with increasing $g$~\cite{bulgac, machida, sensarma}.
This is because the energy separation between the bound states
increases with increasing $|\Delta_0|$ which makes them less occupied.
To further understand these peculiar density depletions, next we analyze the single particle 
density of states for $\sigma$ fermions as well as the spectrum of energy eigenvalues.

%
%
At $T = 0$ the local single particle density of states for $\sigma$ fermions is
defined by
$
D_\sigma(\mathbf{r},\omega) = -(1/\pi){\rm Im} [\lim_{\varepsilon \to 0^+} G_{\sigma,\sigma}(\mathbf{r},\mathbf{r};i\omega_\ell \to \omega + i\varepsilon)].
$
This leads to 
$
D_\sigma(\mathbf{r},\omega) = \sum_{n} [ 
|u_{n,\sigma}(\mathbf{r})|^2 \delta(\omega - \epsilon_{n,\sigma}) 
+ |v_{n,\sigma}(\mathbf{r})|^2 \delta(\omega + \epsilon_{n,-\sigma}) 
]
$
where $\delta(x)$ is the delta function, and it can be written as
$
D_\uparrow(\mathbf{r},\omega) = \sum_{n}|u_n(\mathbf{r})|^2\delta(\omega - \epsilon_n)
$
and
$
D_\downarrow(\mathbf{r},\omega) = \sum_{n} |v_n(\mathbf{r})|^2\delta(\omega + \epsilon_n)
$
by using the symmetry of the BdG equations. 
Then the overall single particle density of states is found by 
$
D_\sigma(\omega) = \int d\mathbf{r} D_\sigma(\mathbf{r},\omega),
$
which for a single vortex reduces to
$
D_\uparrow(\omega) = \sum_{n,m,j} c_{n,j}^2 \delta(\omega - \epsilon_n)
$
and
$
D_\downarrow(\omega) = \sum_{n,m,j} d_{n,j}^2 \delta(\omega + \epsilon_n).
$
We use a small spectral broadening ($0.01\epsilon_F$) to regularize
these delta functions.

\begin{figure} [htb]
\centerline{\scalebox{0.5}{\includegraphics{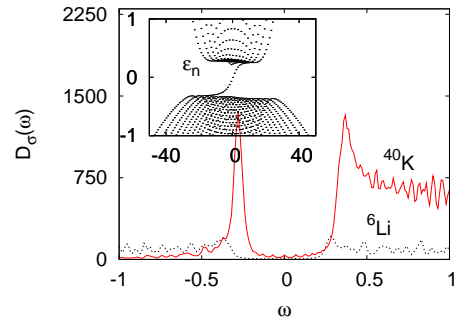} }}
\caption{\label{fig:dos}
(Color online) Single particle density of states for $\sigma$ fermions $D_\sigma(\omega)$ 
(in units of $1/\epsilon_F$) versus energy $\omega$ (in units of $\epsilon_F$) is 
shown for a population balanced mixture of $^6$Li and $^{40}$K atoms. 
Here $|\epsilon_b| = 0.1\epsilon_F$. The inset shows the spectrum of energy eigenvalues 
$\epsilon_n$ (in units of $\epsilon_F$) versus angular momentum $m$ for the same parameters.
}
\end{figure}

In Fig.~\ref{fig:dos} we show $D_\sigma(\omega)$ for a population balanced 
mixture of $^6$Li and $^{40}$K atoms. We also show the spectrum of energy 
eigenvalues $\epsilon_n$ as an inset for the same parameters. Similar to mass 
balanced mixtures~\cite{gygi, nygaard, bulgac, machida, sensarma}, the positive and 
negative energy spectra are connected by a single branch of discrete Andreev-like bound states. 
The visible discreteness of the continuum spectrum shown in Fig.~\ref{fig:dos} is a 
finize size effect and the spectrum becomes continuous only in the thermodynamic limit 
($k_F R \to \infty$), while the discreteness of the bound states is insensitive 
to the system size since these states are strongly localized around the vortex core.
However the energy spectrum is asymmetric around $\epsilon_n = 0$ with less bound 
and continuum states for $\epsilon_n > 0$, which is due to the broken pseudospin 
symmetry since the masses ($m_{K} > m_{Li}$) and therefore the chemical 
potentials ($\mu_{Li} > \mu_{K}$) are different for $\uparrow$ and $\downarrow$ 
fermions. This is in sharp contrast with mass and population balanced mixtures 
where the energy spectrum is symmetric~\cite{machida, sensarma}.

Furthermore the energy spectrum is qualitatively different on the positive and 
the negative $\omega$ sides, and it is very illustrative to make an analogy between the 
energy spectrum of mass imbalanced mixtures shown in Fig.~\ref{fig:dos} 
and that of the mass balanced mixtures~\cite{machida, sensarma}. For mass and 
population balanced mixtures energy spectra that are qualitatively similar to the 
negative (positive) side with many (few) bound and continuum states occur for small 
(large) values of $|\epsilon_b|$, leading to low (high) density depletions at the 
vortex core. This analogy suggests that the local vortex core density of the heavy-mass 
fermions should deplete more than that of the light-mass fermions since the density 
of states for heavy-mass fermions is higher (lower) for positive (negative) $\omega$. 
We also find that the bound state contribution to $n_\sigma(r)$ is  a nonmonotonic 
function of $r$ with a maximum at an intermediate distance $r = r_*$. 
This nonmonotonic contribution is due to the strongly localized 
quasiparticle amplitudes that are associated with the bound states, which also give 
rise to Friedel-like $\Delta(r)$ oscillations around the vortex core in the strict 
BCS limit~\cite{gygi, nygaard, bulgac, machida}.
Therefore the unequal density depletions shown in Fig.~\ref{fig:density} are purely
density of states effects arising from the asymmetric energy spectrum
shown in Fig.~\ref{fig:dos}.
Next we analyze the density as well as the velocity of the superfluid fermions.

%
%
The quantum mechanical probability current operator for $\sigma$ fermions is given by
$
\widehat{\mathbf{J}}_\sigma (\mathbf{r}) = [1/(2M_\sigma i)]  
[\psi_\sigma^\dagger(\mathbf{r}) \nabla \psi_\sigma(\mathbf{r}) - H.c.]
$
where $H.c.$ is the Hermitian conjugate. 
Therefore the local current density 
$
\mathbf{J}_\sigma (\mathbf{r})  = \langle \widehat{\mathbf{J}}_\sigma (\mathbf{r}) \rangle
$ 
circulating around a single vortex becomes
$
\mathbf{J}_\uparrow (\mathbf{r}) = [1/(2M_\uparrow i)] \sum_{n}
[u_n^*(\mathbf{r}) \nabla u_n(\mathbf{r}) f(\epsilon_n) - H.c.]
$
for the $\uparrow$ and
$
\mathbf{J}_\downarrow (\mathbf{r}) = [1/(2M_\downarrow i)] \sum_{n}
[v_n(\mathbf{r}) \nabla v_n^*(\mathbf{r}) f(-\epsilon_n) - H.c.]
$
for the $\downarrow$ fermions where we used the symmetry of the BdG equations.
These relations can be written as 
$
\mathbf{J}_\sigma(\mathbf{r}) = n_0(\mathbf{r}) \mathbf{v}_\sigma (\mathbf{r})/2,
$
where $n_0(\mathbf{r})$ is the local superfluid density and 
$
\mathbf{v}_\sigma(\mathbf{r}) = \kappa \widehat{\mathbf{\theta}}/(2M_\sigma r)
$ 
is the local superfluid velocity. Therefore $\mathbf{J}_\sigma(\mathbf{r})$ is along the 
$\widehat{\mathbf{\theta}}$ direction, and for a single vortex it is given by
$
J_\uparrow(r) = [1/(2\pi M_\uparrow r)] \sum_{n,m}
m[\sum_j c_{n,j} \phi_{j,m}(r)]^2 f(\epsilon_n)
$
for the $\uparrow$ and
$
J_\downarrow(r) = -[1/(2\pi M_\downarrow r)] \sum_{n,m}
(m+\kappa) [\sum_j d_{n,j} \phi_{j,m+\kappa}(r)]^2 f(-\epsilon_n)
$
for the $\downarrow$ fermions such that $n_0(r) = (4M_\sigma/\kappa) r J_\sigma(r)$.

\begin{figure} [htb]
\centerline{\scalebox{0.5}{\includegraphics{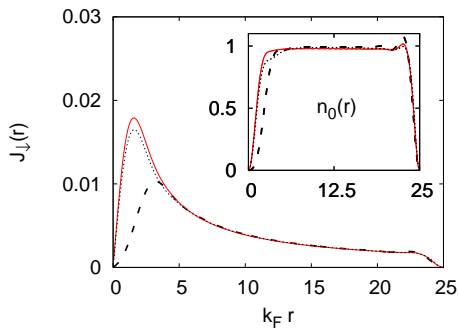} }}
\caption{\label{fig:current}
(Color online) Local current density $J_\downarrow(r)$ [in units of $M_\downarrow/k_F^3$] versus 
radius $r$ (in units of $1/k_F$) is shown for a population balanced mixture 
of $^6$Li and $^{40}$K atoms. Here $|\epsilon_b| = 0.1\epsilon_F$ 
(dashed line), $|\epsilon_b| = 0.3\epsilon_F$ (solid line) and 
$|\epsilon_b| = 0.5\epsilon_F$ (dotted line). The inset shows the local superfluid fermion 
density $n_0(r)$ [in units of $k_F^2/(2\pi)$] versus $r$ for the same parameters.
}
\end{figure}

In Fig.~\ref{fig:current} we show $J_\downarrow(r)$ for a population balanced 
mixture of $^6$Li and $^{40}$K atoms. We also show $n_0(r)$ as an inset 
for the same parameters. The bound states have positive (paramagnetic) and the 
continuum states have negative (diamagnetic) contribution to $J_\downarrow(r)$.
This leads to a nonmonotonic $J_\downarrow(r)$ which first increases as $\propto r$ 
and then decreases as $\propto 1/r$. The latter behavior is due to
the saturation of $n_0(r)$ for long distances away from the vortex core. 
Therefore a maximum peak current occurs for all values of $|\epsilon_b|$ 
at some distance $r_c$ away from the vortex core. However the value of this peak 
current increases until $|\epsilon_b| \simeq 0.3\epsilon_F$ and then decreases for 
higher values of $|\epsilon_b|$. 
Since a two-body bound state exists even for an arbitrarilly small $g > 0^+$ in two 
dimensions, we emphasize that this nonmonotonic evolution is not due to
the occurrence of a two-body bound state threshold
(divergence of the two-body scattering length), as previously suggested 
for mass and population balanced mixtures in three dimensions~\cite{sensarma}. 
We believe that it is related to the nonmonotonic evolution of the coherence length 
$\xi_c$ which can be easily extracted from $n_0(r)$. 
This is qualitatively consistent with the recent experiments involving mass and 
population balanced mixture of $^6$Li atoms, where a pronounced peak of 
critical velocity has been observed on the molecular side of the strongly 
interacting regime in one-dimensional optical lattices~\cite{miller}.

%
%
In summary we analyzed the vortex core states of population balanced $^{6}$Li and $^{40}$K 
mixture at $T = 0$ as a function of two-body binding energy. We found that the vortex core 
is mostly occupied by the light-mass ($^{6}$Li) fermions and that the core density of 
the heavy-mass ($^{40}$K) fermions is highly depleted. 
This is in contrast with mass and population balanced mixtures where an equal amount of 
density depletion is found at the vortex core for both pseudospin components. 

This work was partially completed in Georgia Tech, and the author gratefully acknowledges 
discussions with C. A. R. S{\'a} de Melo and P. S. Julienne.

\end{document}